\documentclass{appolb}
\usepackage{graphicx,url}
\begin{document}
\title{Measuring dilepton and heavy quark production\\ at large $\mu_{\rm B}$: the NA60+ experiment at the CERN SPS %
\thanks{Presented at the 29$^{\rm th}$ International Conference on Ultrarelativistic Nucleus Nucleus Collisions, April 4-10 2022, Krak\'ow, Poland}%
}
\author{Enrico Scomparin, for the NA60+ Collaboration
\address{INFN Torino, Italy}
%\\[3mm]
%{Third Author % of different affiliation
%\address{affiliation}
}%
%\\[3mm]
%the Name(s) of other Author(s)
%\address{affiliation}
%}
\maketitle
\begin{abstract}
The high-$\mu_{\rm B}$ region of the QCD phase diagram has become the object of several studies, focused on the investigation of the order of the phase transition and the search for the critical point. Accessing rare probes
%, which include electromagnetic observables and heavy quark production, 
is experimentally challenging as it requires large integrated luminosities, and a fixed-target environment represents an ideal solution for these studies. The CERN SPS covers, with large beam intensity, the collision energy region $5<\sqrt{s_{\rm NN}}<17$ GeV.  
%which was little studied until now with rare observables. 
A future experiment, NA60+, is being proposed to access this region and perform accurate measurements of the dimuon spectrum from threshold up to the charmonium mass region, and of hadronic decays of charm and strange hadrons. 
%via their 2- and 3-body hadronic decays. 
The experiment, which is also part of the Physics Beyond Colliders CERN initiative, 
%includes a muon spectrometer, based on tracking gas detectors (GEM, MWPC) coupled to a vertex spectrometer based on Si detectors (MAPS). 
aims at taking its first data with Pb and proton beams around 2029.  
\end{abstract}
  
\section{Introduction}
Studies of electromagnetic and hard probes have represented one of the most important sources of information on the early stages of the strongly interacting partonic system created in ultra-relativistic heavy-ion collisions. 
Virtual photons, which can be detected as lepton pairs, are emitted all along the collision history but their yield is dominated by the high-temperature phase. The $T_{\rm slope}$ parameter of the spectral shape of the dilepton invariant mass spectrum can be considered as a space-time average of the thermal temperature $T$ over the  fireball evolution. Measurements carried out by the HADES and NA60 experiments, respectively in Au--Au collisions at $\sqrt{s_{\rm NN}} = 2.42$ GeV and In--In at $\sqrt{s_{\rm NN}} = 17.3$ GeV, allowed experimental estimates of the medium temperature. Values increasing from $T_{\rm slope} = 71.8\pm2.1$ MeV to $205 \pm 12$ MeV were obtained, showing that the crossing of the pseudo-critical temperature should be reached in the collision energy interval defined by the two existing measurements. In such a range, measurements of lepton pairs allow a study of modifications of the spectral function of the $\rho_0$ and of its chiral partner $a_{\rm 1}$, related to chiral symmetry restoration. The latter resonance is not directly coupled to the dilepton channel, but its modification leads to an enhancement of the continuum in the region $1<m_{\rm ll}<1.4$ GeV/c$^2$. Precise measurements of the dilepton observables outlined in this paragraph are missing below top SPS energy ($\sqrt{s_{\rm NN}} = 17.3$ GeV) and represent a solid motivation for a dedicated experiment at this facility, to scan the range $5<\sqrt{s_{\rm NN}}<17$ GeV. 

Such a dilepton experiment can also access quarkonium production. A measurement of the J/$\psi$ suppression, and possibly of excited states ($\psi({\rm 2S})$, $\chi_{\rm c}$) is of utmost importance, seen its relation with deconfinement, studied in great detail from top SPS to RHIC and LHC energy. In addition to hidden charm, open charm production represents a crucial measurement to access information on the transport coefficients of the medium, as well as its hadronization mechanisms. 
%Open charm contributes to the dilepton mass spectrum continuum, through the semi-leptonic decay of charm hadron pairs. 
Accurate measurements of open charm hadrons can only be performed by accessing their hadronic decays, and therefore a complete study of heavy quark production at SPS energy requires the capability of tracking charged particles in addition to dilepton measurements.

This situation has led to studies for NA60+, a new experiment at the CERN SPS~\cite{Dahms:2673280} based on a muon spectrometer coupled to a vertex spectrometer, and inspired to the former NA60 experiment. In the next sections we will give details on the foreseen experimental set-up and related detector studies, and will briefly recall some estimates of the physics performance and finally the planning for the preparation of the experiment.

\section{Experimental apparatus and detector studies}

\begin{figure}[htb]
\centerline{%
\includegraphics[width=12.5cm]{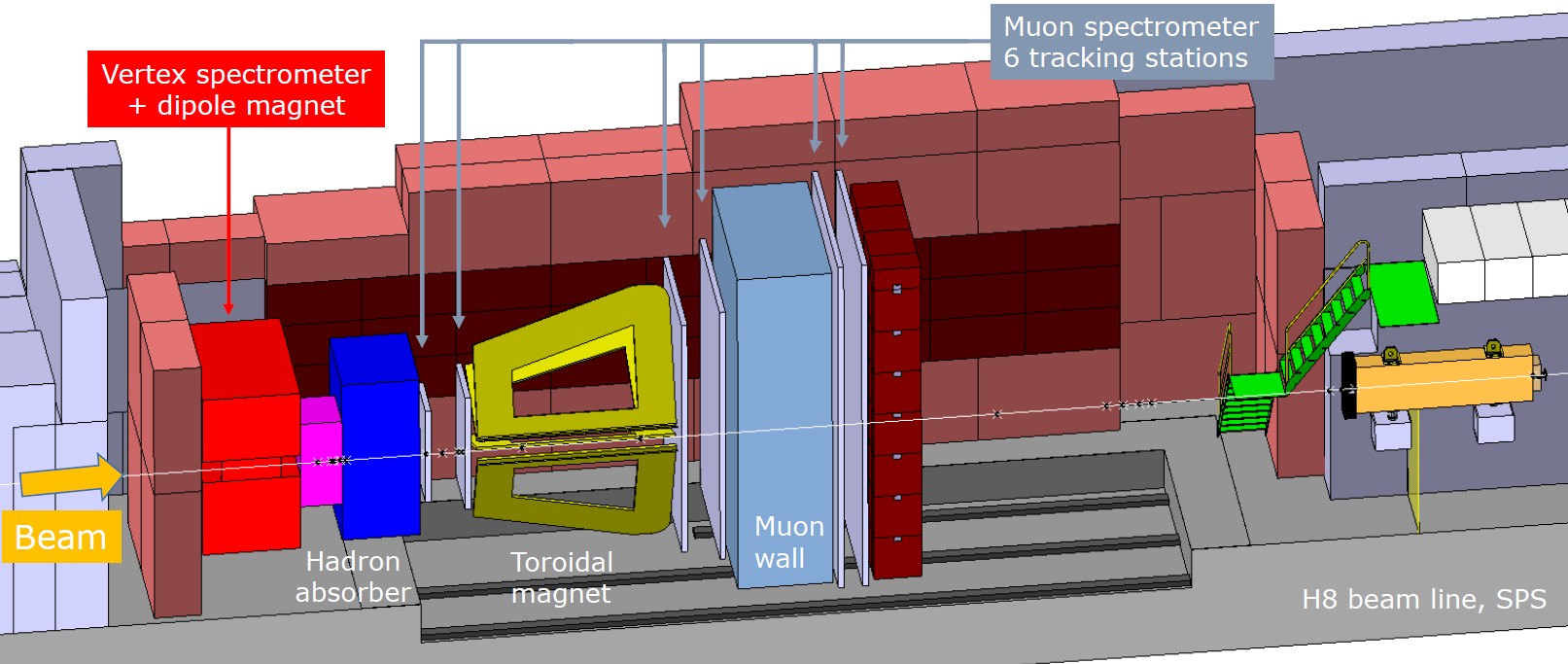}}
\caption{Sketch of the experimental set-up (top view). See text for details.}
\label{Fig:setup}
\end{figure}

A sketch of the experiment can be seen in Fig.~\ref{Fig:setup}. The muon spectrometer is separated from the target region by means of a thick hadron absorber, made of BeO and graphite that filters out the hadrons. Candidate muons are tracked by means of six tracking stations. The two upstream stations are followed by a large-aperture toroidal magnet and by two further tracking stations. A ``muon wall'' acts as a further filter to remove hadrons escaping the absorber as well as low-momentum muons and is followed by the two final tracking stations. In the target region a system of five closely spaced targets is followed by a vertex spectrometer which includes from five up to ten tracking MAPS planes. The vertex spectrometer is embedded in the gap of a dipole magnet. By matching muon tracks in the muon and vertex spectrometers, in the coordinate and momentum space, a  measurement of the muon kinematics not affected by multiple scattering and energy loss in the hadron absorber becomes possible. The muon spectrometer is mounted on a system of rails that allows keeping a constant rapidity coverage when changing the collision energy. When moving the spectrometer downstream along the beam line (high energy collisions) the thickness of the absorber will be increased by adding further graphite elements.
%\subsection{Subsection}
%The text...

\begin{figure}[htb]
\centerline{%
\includegraphics[width=6.35cm]{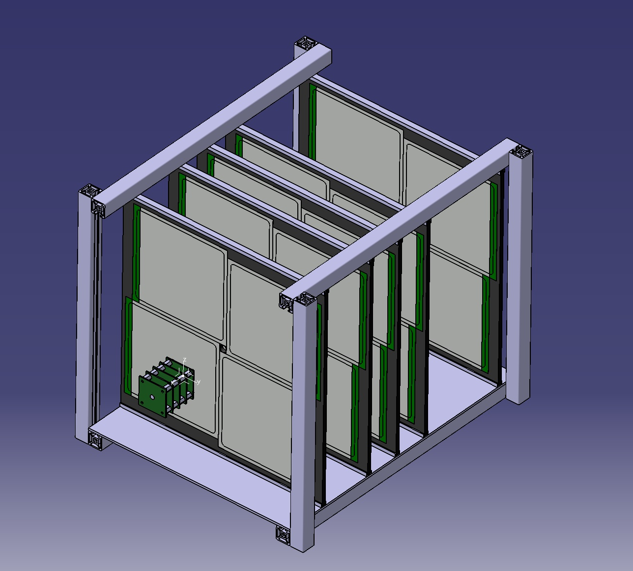}
\includegraphics[width=5.65cm]{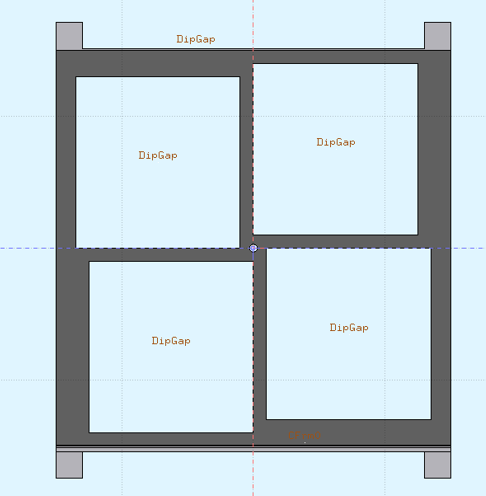}}
\caption{Possible layout of the target system and the vertex spectrometer (left); sketch of the mechanical support of a MAPS plane (right).}
\label{Fig:target}
\end{figure}

In Fig.~\ref{Fig:target} (left) a possible arrangement of the vertex spectrometer is shown. Each plane contains four large MAPS sensors ($15\times15$ cm$^2$ each). The R\&D phase is in progress in the frame of a common development of the ALICE (ITS3 project) and NA60+ collaborations. For NA60+ each sensor is based on 25 mm long units, replicated several times through a stitching procedure. For each stitched long sensor, the control logic to steer the priority encoders, the interfaces for the configuration of the chip and serial data transmitters are all located at the periphery of the sensor, outside the detector acceptance. The thickness of the sensor will be $<0.1$\% radiation length and the spatial resolution will be $\le 5$ $\mu$m. From the mechanical point of view the sensors will be mounted on a light frame of carbon fiber or foam, as shown in Fig.~\ref{Fig:target}(right). 
%Preliminary studies of cooling options show that a mixed fluid + air system could allow an efficient control of the temperature over all the surface of the detection plane.

\begin{figure}[htb]
\centerline{%
\includegraphics[width=8cm]{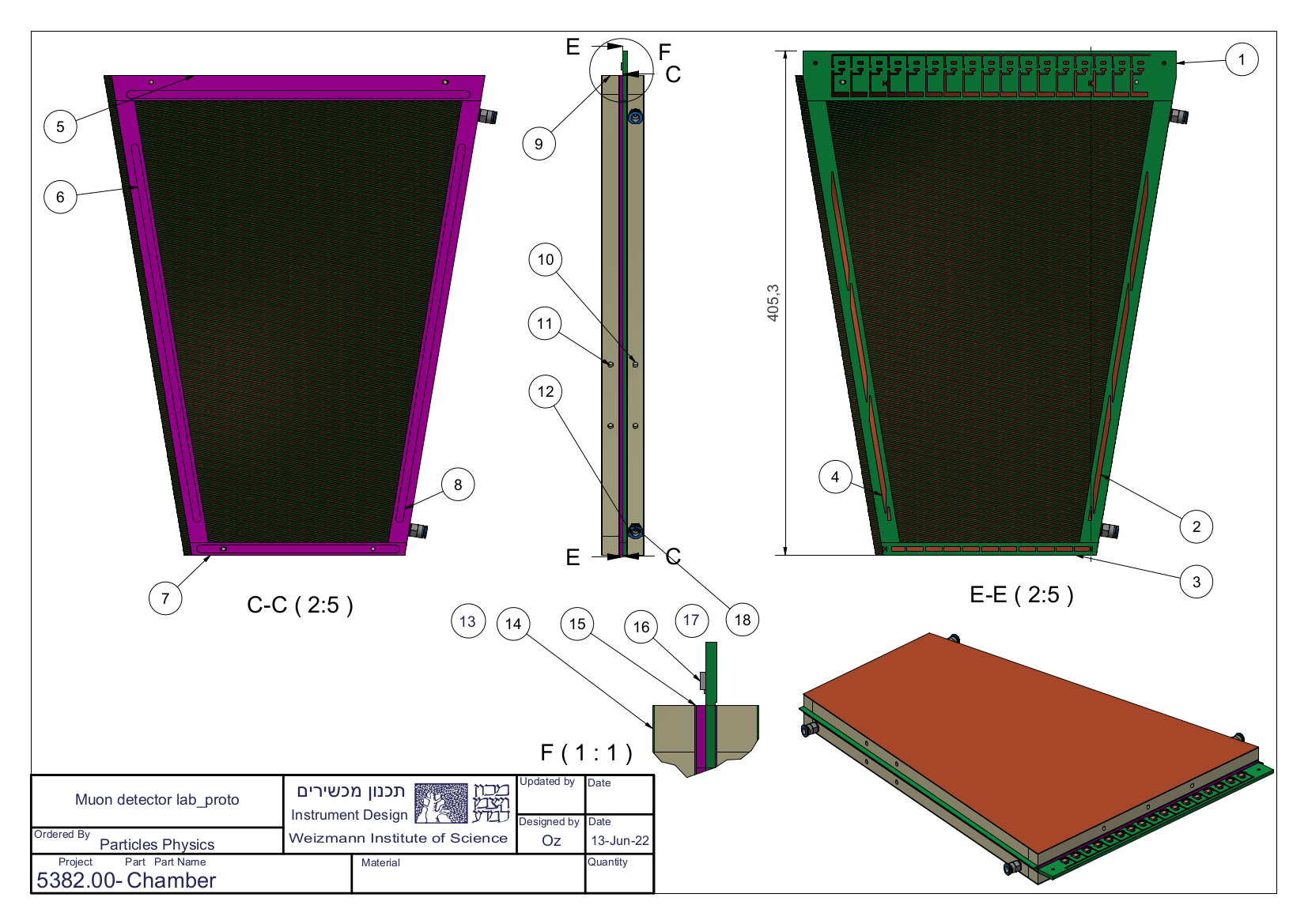}
\includegraphics[width=4cm]{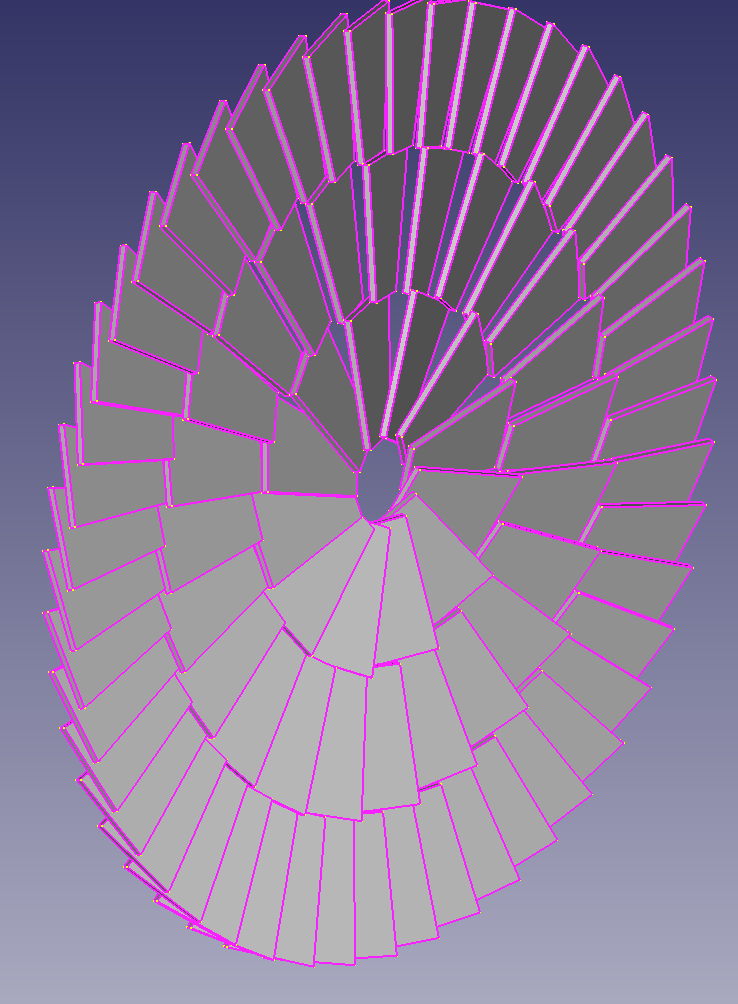}}
\caption{Design of the first prototype of MWPC unit for the NA60+ muon spectrometer (left); layout of the largest muon tracking stations
%, downstream of the Muon wall 
(right).}
\label{Fig:muonspectro}
\end{figure}

For the muon tracking, due to the relatively low rates after the hadron absorber ($\sim 2$ kHz, at the expected Pb beam rate of $10^6$ s$^{-1}$) GEMs or MWPCs options are considered. For the latter, a prototype is being built and tested in the lab, and will be exposed to an SPS test beam. The tracking stations will be built by replicating the prototype detector in such a way to obtain the desired geometry. The two upstream stations will include 12 trapezoid detectors each, while those downstream of the magnet and the muon wall will have 36 and 84 detectors respectively. Figure~\ref{Fig:muonspectro} shows the design of the MWPC unit (left) and the foreseen arrangement for the largest stations (right). The GEM option will also be studied, based on the triple GEM detector studied for the MOLLER experiment at JLab.

\begin{figure}[htb]
\centerline{%
\includegraphics[width=7cm]{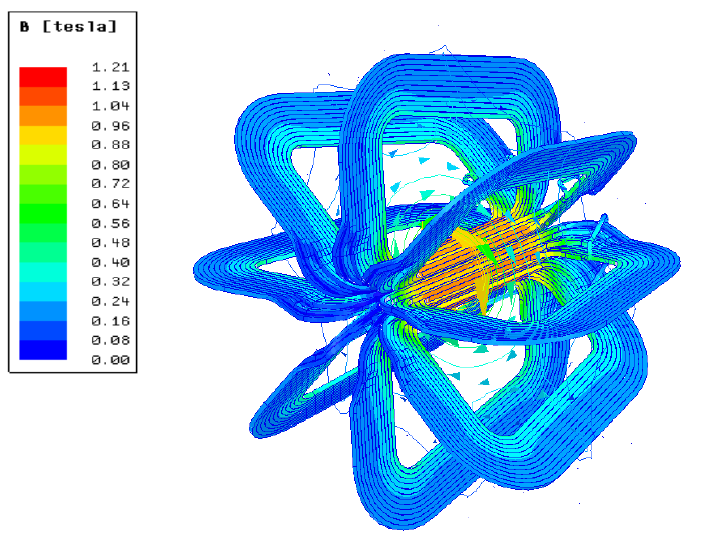}
\includegraphics[width=5cm]{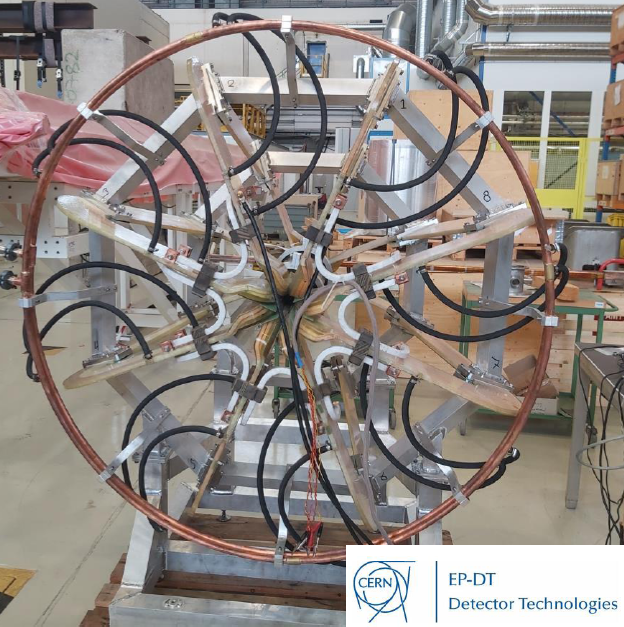}}
\caption{Magnetic field calculations for the NA60+ toroidal magnet (left); the scale 1:5 demonstrator under test (right).}
\label{Fig:magnet}
\end{figure}

The toroidal magnet also represents a major element of the set-up. A warm pulsed magnet is foreseen, producing 0.5 T of magnetic field magnitude over a volume of 120 m$^3$.
The toroid is made from eight sectors, each coil has 12 turns and the conductor has a square copper section with 50 mm side length and a circular cooling channel in the centre. The current is 190 kA, and 
%and all coils are connected in series, giving 
the total power is 3 MW. 
%To test the technology aspects,
%needed to wind the large normal-conducting coil, 
A demonstrator (scale 1:5) was constructed and tested, allowing a cross-check of various aspects of the design. Measurements of the magnetic field in the prototype were found to be in agreement with simulations within 3\%. In Fig.~\ref{Fig:magnet} (left) the results of the field calculations for the toroidal magnet are reported, while on the right panel a photo of the demonstrator can be seen. Finally, for the dipole magnet we plan to use the MEP48 magnet, currently stored at CERN. It can deliver a 1.5 T field, over a 40 cm gap. 

After integration studies, the PPE138 hall on the H8 beam line of the SPS was chosen as installation site for the NA60+ set-up. Radio-protection calculations have led to the design of an iron/concrete shielding, to keep the dose levels below 3 $\mu$Sv/h externally to the experimental hall for the intended 10$^6$ s$^{-1}$ Pb beam intensity. Specific beam optics studies have shown that a sub-millimetric beam can be delivered to the experiment from $E=30$ to 158 AGeV.   

\section{Physics performance}

\begin{figure}[htb]
\centerline{%
\includegraphics[width=6cm]{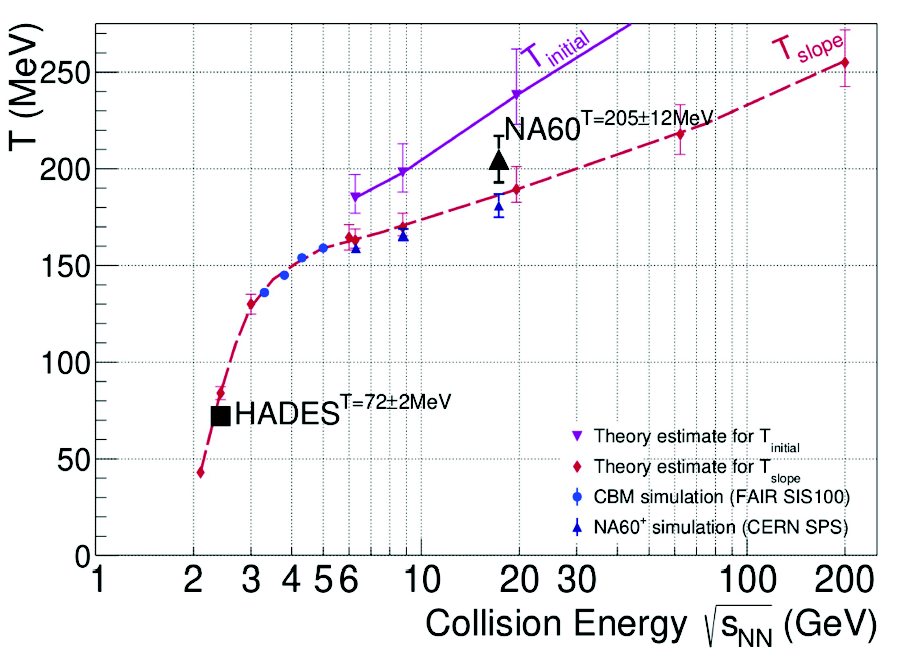}
\includegraphics[width=7cm]{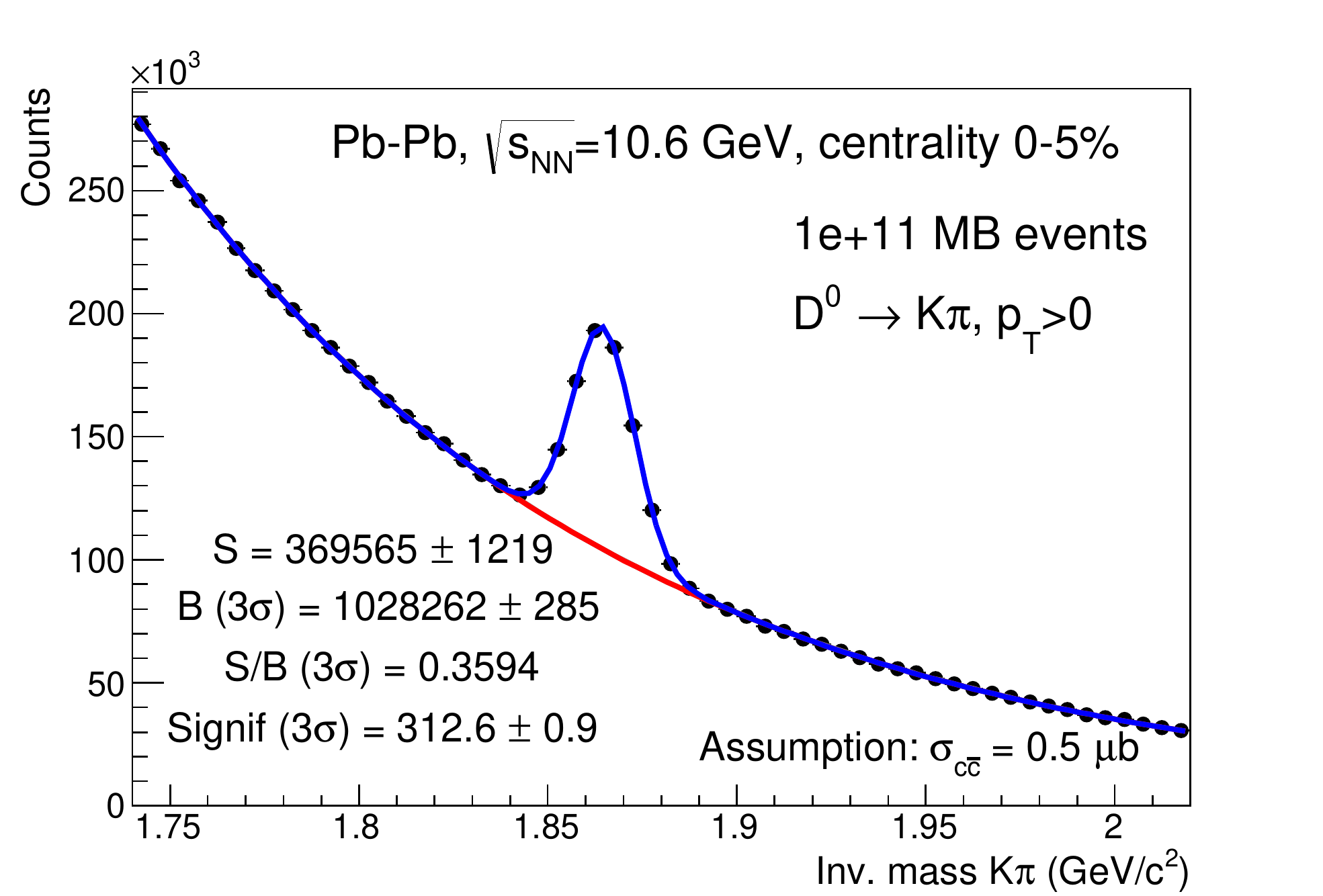}}
\caption{Energy dependence of the $T_{\rm slope}$ parameter. The blue triangles show the expected NA60+ performance (left); invariant mass distribution of candidates for the $D^{\rm 0}\rightarrow{\rm K}\pi$ decay at $E=60$ AGeV (right).}
\label{Fig:physics}
\end{figure}

Some of the main results of performance studies are briefly recalled here. With the expected Pb beam intensity, we expect to collect 10$^{11}$ minimum bias collision events in the typical 1-month period dedicated to heavy-ion running at CERN. The corresponding dimuon statistics for central (0--5\%) events amounts to $\sim 4\cdot 10^6$. For the measurement of the $T_{\rm slope}$ parameter, uncertainties as low as $\sim 3\%$ can be reached. The result on the measurement of the energy dependence of $T_{\rm slope}$ is  shown in Fig.~\ref{Fig:physics} (left) together with existing measurements from HADES and NA60 and corresponding studies for the CBM experiment at GSI, which explores a lower, complementary energy range. It will be possible to explore the region close to the pseudocritical temperature and detect posssible signals of a first-order phase transition that can be expected at these energies. 

Open charm measurements will be performed starting from the tracks reconstructed in the vertex spectrometer. Charm hadron candidates are built by combining pairs or triplets of tracks with the proper charge signs. The huge combinatorial background can be reduced via geometrical selections on the displaced decay-vertex topology. 
%exploiting the fact that the mean proper
%decay lengths $c\tau$ are of about 60–310 $\mu$m, and therefore their decay vertices
%are typically displaced by a few hundred $\mu$m from the primary interaction vertex. 
In Fig.~\ref{Fig:physics} (right) the expected signal from the $D^{\rm 0}\rightarrow{\rm K}\pi$ decay, for $E= 60$ AGeV incident energy, is shown. Measurements of hidden charm production will  also be performed. The expected statistics range from $10^4$ to $10^5$ J/$\psi$ from low to high SPS energy, allowing precise study of the suppression signal. Corresponding measurements in p-A collisions will also be carried out.
%in order to evaluate the contribution of cold nuclear matter effects. 

\section{Planning and conclusions}

After the submission of an EoI in 2019~\cite{Dahms:2673280}, the Letter of Intent will be sent to the CERN SPSC at the end of 2022. The current plan, if the experiment will be approved and funded, is to be able to take the first data in 2029, after the currently foreseen end of the LHC Shutdown 3. The running plan for the first 5-6 years includes a $\sim 1$ month period with a Pb beam, with a different incident energy for each year, complemented by a few weeks of proton beam running for calibration purposes and physics studies with p-A collisions. 

\bibliographystyle{utphys}
\footnotesize
\bibliography{main}
%\bibliography{main}

\end{document}